\pdfoutput=1
\documentclass[aps,prb,twocolumn,noshowpacs,a4paper]{revtex4-1}

\usepackage{graphicx}
\usepackage{color}
\usepackage{textcomp}
\usepackage{amssymb}\usepackage{amsmath}\usepackage{amsthm}
\usepackage{units}

\begin{document}

\title{The Dicke Quantum Phase Transition with a Superfluid Gas in an Optical Cavity}

\author{Kristian Baumann}
\affiliation{Institute for Quantum Electronics, ETH Z\"{u}rich,
CH--8093 Z\"{u}rich, Switzerland}
\author{Christine Guerlin}
\altaffiliation[Present adresse:]{Thales Research and Technology,
Campus Polytechnique, 1 Avenue Augustin Fresnel, F-91767
Palaiseau, France} \affiliation{Institute for Quantum Electronics,
ETH Z\"{u}rich, CH--8093 Z\"{u}rich, Switzerland}
\author{Ferdinand Brennecke}
\affiliation{Institute for Quantum Electronics, ETH Z\"{u}rich,
CH--8093 Z\"{u}rich, Switzerland}
\author{Tilman Esslinger}
\email{Email: esslinger@phys.ethz.ch} \affiliation{Institute for
Quantum Electronics, ETH Z\"{u}rich, CH--8093 Z\"{u}rich,
Switzerland}

\date{\today}

\begin{abstract}
A phase transition describes the sudden change of state in a physical system, such as the transition between a fluid and a solid. Quantum gases provide the opportunity to establish a direct link between experiment and generic models which capture the underlying physics. A fundamental concept to describe the collective matter-light interaction is the Dicke model which has been predicted to show an intriguing quantum phase transition. Here we realize the Dicke quantum phase transition in an open system formed by a Bose-Einstein condensate coupled to an optical cavity, and observe the emergence of a self-organized supersolid phase. The phase transition is driven by infinitely long-range interactions between the condensed atoms. These are induced by two-photon processes involving the cavity mode and a pump field. We show that the phase transition is described by the Dicke Hamiltonian, including counter-rotating coupling terms, and that the supersolid phase is associated with a spontaneously broken spatial symmetry. The boundary of the phase transition is mapped out in quantitative agreement with the Dicke model. The work opens the field of quantum gases with long-range interactions, and provides access to novel quantum phases.
\end{abstract}
\maketitle

\section*{Introduction}
The realization of Bose-Einstein condensation (BEC) in a dilute atomic gas\cite{anderson1995, davis1995c} marked the beginning of a new approach to quantum many-body physics. Meanwhile, quantum degenerate atoms are regarded as an ideal tool to study many-body quantum systems in a very well controlled way. Excellent examples are the BEC-BCS crossover\cite{regal2004, zwierlein2004, bartenstein2004} and the observation of the superfluid to Mott-insulator transition\cite{greiner2002}. The high control available over these many-body systems has also stimulated the notion of quantum simulation\cite{feynman1982,lloyd1996}, one of the goals being to generate a phase diagram of an underlying Hamiltonian. However, the phase transitions and crossovers which have been experimentally investigated with quantum gases up to now are conceptually similar since their physics is governed by short-range interactions.

In order to create many-body phases dominated by long-range interactions different routes have been suggested, most of which exploit dipolar forces between atoms and molecules\cite{lahaye2009}. A rather unique approach considers atoms inside a high-finesse optical cavity, so that the cavity field mediates infinitely long-range forces between all atoms\cite{asboth2004,asboth2007}. In such a setting a phase transition from a Bose-Einstein condensate to a self-organized phase has been predicted once the atoms induce a sufficiently strong coupling between a pump field and an empty cavity mode\cite{domokos2002,nagy2008}. Indeed, self-organization of a classical, laser-cooled atomic gas in an optical cavity was observed experimentally\cite{black2003}. Conceptually related experiments studied the atom-induced coupling between a pump field and a vacuum mode using ultracold or condensed atoms. This led to the observation of free-space\cite{inouye1999,yoshikawa2005} and cavity-enhanced\cite{slama2007} superradiant Rayleigh scattering, as well as to collective atomic recoil lasing\cite{slama2007,bonifacio1994}. Both phenomena did not support steady-state quantum phases, and became visible in transient matter wave pulses.

A rather general objective of many-body physics is to understand quantum phase transitions\cite{sachdev1999} and to unravel their connection to entanglement\cite{amico2008, osterloh2002}. An important concept within this effort is a system of interacting spins in which each element is coupled to all others with equal strength. The most famous example for such an infinitely coordinated\cite{botet1982} spin system is the Dicke model\cite{dicke1954}, which has been predicted to exhibit a quantum phase transition more than thirty years ago\cite{hepp1973, wang1973}. The Dicke model considers an ensemble of two-level atoms, i.e. spin-$1/2$ particles, coupled to a single electromagnetic field mode. For sufficient coupling this system enters a superradiant phase with macroscopic occupation of the field mode. A promising route to realize this transition has been proposed recently in the setting of cavity quantum electrodynamics by Carmichael and coworkers\cite{dimer2007}. In their scheme strong coupling between two ground states of an atomic ensemble is induced by balanced Raman transitions involving a cavity mode and a pump field. This idea circumvents the thought to be unattainable condition for the Dicke quantum phase transition which requires a coupling strength on the order of the energy separation between the two involved atomic levels.

In this work we realize the Dicke quantum phase transition in an open system and observe self-organization of a Bose-Einstein condensate. In the experiment, a condensate is trapped inside an ultrahigh-finesse optical cavity, and pumped from a direction transverse to the cavity axis, as shown in figure 1. We will theoretically show that the onset of self-organization is equivalent to the Dicke quantum phase transition where the two-level system is formed by two different momentum states which are coupled via the cavity field. At the phase transition a spatial symmetry of the underlying lattice structure, given by the pump and cavity modes, is spontaneously broken. This steers the system from a flat superfluid phase into a quantum phase with macroscopic occupation of the higher-order momentum mode and the cavity mode. The corresponding density wave together with the presence of off-diagonal long-range order allows to regard the organized phase as a supersolid\cite{andreev1969,chester1970, leggett1970} similar to those proposed for two-component systems\cite{buchler2003}.
\begin{figure}
\centering
\includegraphics{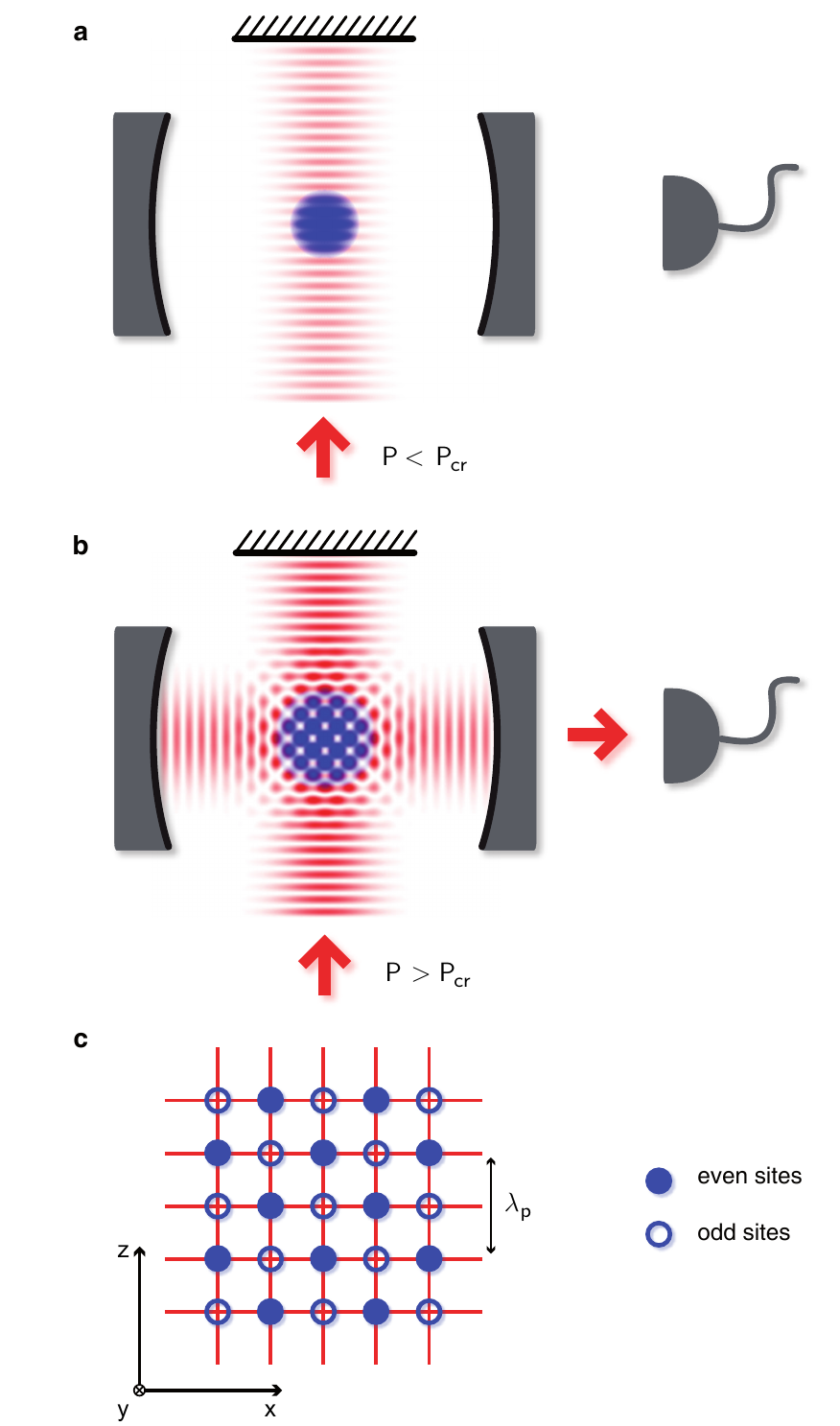}
\caption{Concept of the experiment. A Bose-Einstein condensate which is placed inside an optical cavity is driven by a standing-wave pump laser oriented along the vertical $z$-axis. The frequency of the pump laser is far red-detuned with respect to the atomic transition line but close detuned to a particular cavity mode. Correspondingly, the atoms coherently scatter pump light into the cavity mode with a phase depending on their position within the combined pump--cavity mode profile. \textbf{a}, For a homogeneous atomic density distribution along the cavity axis, the build-up of a coherent cavity field is suppressed due to destructive interference of the individual scatterers. \textbf{b}, Above a critical pump power $P_\mathrm{cr}$ the atoms self-organize onto either the even or odd sites of a checkerboard pattern (\textbf{c}) thereby maximizing cooperative scattering into the cavity. This dynamical quantum phase transition is triggered by quantum fluctuations in the condensate density. It is accompanied by spontaneous symmetry breaking both in the atomic density and the relative phase between pump field and cavity field. \textbf{c}, Geometry of the checkerboard pattern. The intensity maxima of the pump and cavity field are depicted by the horizontal and vertical lines, respectively, with $\lambda_p$ denoting the pump wavelength.}
\end{figure}
\section*{Theoretical Description and the Dicke Model}

Let us first consider a single two-level atom of mass $m$ interacting with a single cavity mode and the standing-wave pump field. The Hamiltonian then reads\cite{maschler2008} in a frame rotating with the pump laser frequency
\begin{align}
\hat{H}_{(1)} = \frac{\hat{p}_x^2 + \hat{p}_z^2}{2m} &+ V_0 \cos^2(k\hat{z}) + \hbar\eta (\hat{a}^\dag + \hat{a}) \cos(k\hat{x})\cos(k\hat{z}) \notag\\&- \hbar\Big(\Delta_c - U_0\cos^2(k\hat{x})\Big) \hat{a}^\dag \hat{a}.
\end{align}
Here, the excited atomic state is adiabatically eliminated which is justified for large detuning $\Delta_a = \omega_p - \omega_a$ between the pump laser frequency $\omega_p$ and the atomic transition frequency $\omega_a$. The first term describes the kinetic energy of the atom with momentum operators $\hat{p}_{x,z}$. The pump laser creates a standing-wave potential of depth $V_0 = \hbar \Omega_p^2/\Delta_a$ along the $z$-axis, where $\Omega_p$ denotes the maximum pump Rabi frequency, and $\hbar$ the Planck constant. Scattering between the pump field and the cavity mode, which is oriented along $x$, induces a lattice potential which dynamically depends on the scattering rate and the relative phase between the pump field and the cavity field. This phase is restricted to the values 0 or $\pi$, for which the scattering induced light potential has a $\lambda_p/\sqrt{2}$ periodicity along the $x$-$z$ direction, with $\lambda_p = 2\pi/k$ denoting the pump wavelength (see Fig.~1c). The scattering rate is determined by the two-photon Rabi frequency $\eta = g_0 \Omega_p/\Delta_a$, with $g_0$ being the atom-cavity coupling strength. The last term describes the cavity field, with photon creation and annihilation operators $\hat{a}^\dag$ and $\hat{a}$. The cavity resonance frequency $\omega_c$ is detuned from the pump laser frequency by $\Delta_c = \omega_p - \omega_c$, and the light-shift of a single maximally coupled atom is given by $U_0 = \frac{g_0^2}{\Delta_a}$.

For a condensate of $N$ atoms, the process of self-organization can be captured by a mean-field description\cite{nagy2008}. It assumes that all atoms occupy a single quantum state characterized by the wave function $\psi$, which is normalized to the atom number $N$. The light-atom interaction can now be described by a dynamic light potential\cite{domokos2003} felt by all atoms. Since the timescale of atomic dynamics in the motional degree of freedom is much larger than the inverse of the cavity field decay rate $\kappa$, the coherent cavity field amplitude $\alpha$ adiabatically follows the atomic density distribution according to $\alpha = \eta\Theta/(\Delta_c - U_0 \mathcal{B} +i\kappa)$. The order parameter describing self-organization is given by $\Theta  = \langle\psi|\cos(kx)\cos(kz)|\psi\rangle$ which measures the localization of the atoms on either the even ($\Theta > 0$) or the odd ($\Theta < 0$) sublattice of the underlying checkerboard pattern defined by $\cos(kx)\cos(kz)=\pm 1$ (see Fig.~1c). The sign of the order parameter determines which of the two possible relative phases is adopted by the cavity field. According to the spatial overlap between the atomic density and the cavity mode profile, the atoms dispersively shift the cavity resonance proportional to $\mathcal{B} = \langle\psi|\cos^2(kx)|\psi\rangle$. The resulting dynamic lattice potential reads
\begin{align}
V(x,z) = & V_0 \cos^2(kz) + \hbar U_0 |\alpha|^2 \cos^2(k x) \notag\\&+ \hbar \eta(\alpha + \alpha^*)\cos(kx) \cos(kz).
\end{align}
The atoms self-organize due to positive feedback from the interference term in equation (2) above a critical two-photon Rabi frequency $\eta_\mathrm{cr}$. Assuming that a density fluctuation of the condensate induces e.g.~$\Theta > 0$, and the pump-cavity detuning is chosen to yield $\Delta_c - U_0\mathcal{B}<0$, the lattice potential resulting from light scattering further attracts the atoms towards the even sites. This in turn increases light scattering into the cavity and starts a runaway process. The system reaches a steady state once the gain in potential energy is balanced by the cost in kinetic energy and collisional energy.
\begin{figure}
\centering
\includegraphics{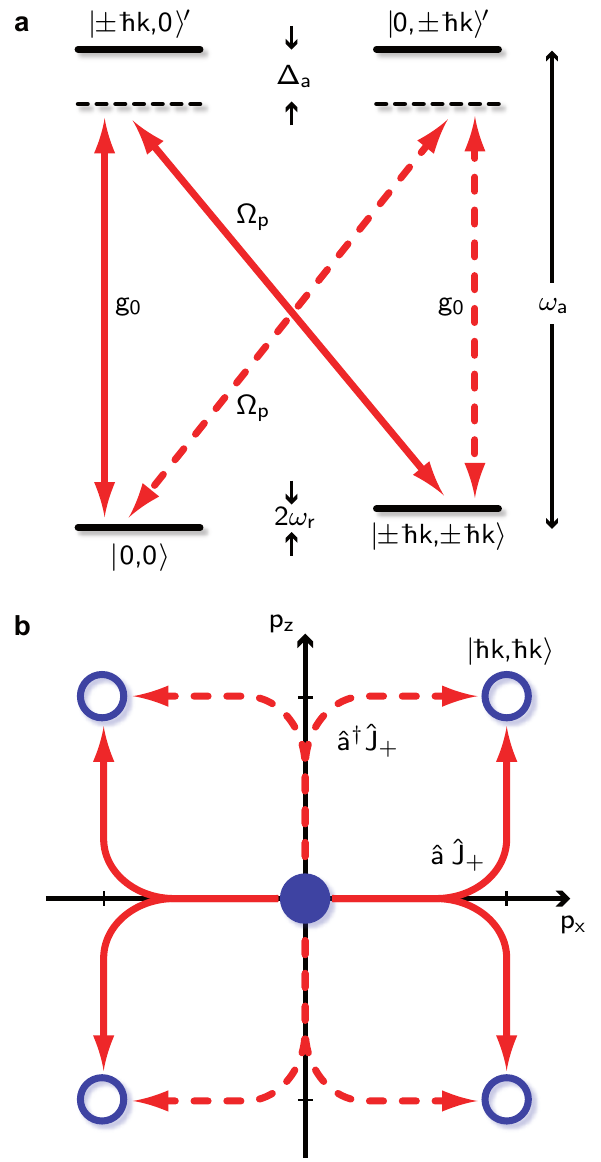}
\caption{Analogy to the Dicke model. In an atomic two-mode picture the pumped BEC--cavity system is equivalent to the Dicke model including counter-rotating interaction terms. \textbf{a}, Light scattering between the pump field and the cavity mode induces two balanced Raman channels between the atomic zero-momentum state $|p_x,p_z\rangle=|0,0\rangle$ and the symmetric superposition of the states $|\pm \hbar k, \pm \hbar k\rangle$ with an additional photon momentum along the $x$ and $z$ directions. \textbf{b}, The two excitation paths (dashed and solid) corresponding to the two Raman channels are illustrated in a momentum diagram. For the notation see text.}
\end{figure}

Fundamental insight into the onset of self-organization is gained from a direct analogy to the Dicke model quantum phase transition\cite{dicke1954, hepp1973, wang1973}. This analogy uses a two-mode description for the atomic field, where the initial Bose-Einstein condensate is approximated by the zero-momentum state $|p_x, p_z\rangle = |0,0\rangle$. Photon scattering between the pump and cavity field couples the zero-momentum state to the symmetric superposition of  states which carry an additional photon momentum along the $x$ and $z$ directions: $|\pm \hbar k, \pm \hbar k \rangle = \sum_{\mu,\nu=\pm 1} |\mu\hbar k, \nu\hbar k\rangle/2$. The energy of this state is correspondingly lifted by twice the recoil energy $E_r = \hbar\omega_r = \hbar^2 k^2/(2m)$ compared to the zero-momentum state. (For the inclusion of Bloch states, see Methods.)

There are two possible paths from the zero-momentum state $|p_x, p_z\rangle = |0,0\rangle$ to the excited momentum state $|\pm \hbar k, \pm \hbar k \rangle$: i) the absorption of a standing-wave pump photon followed by the emission into the cavity, $\hat{a}^\dag \hat{J}_+$, and ii) the absorption of a cavity photon followed by the emission into the pump field, $\hat{a} \hat{J}_+$  (see Fig.~2b). Here, the collective excitations to the higher-energy mode are expressed by the ladder operators $\hat{J}_+ = \sum_i|\pm \hbar k, \pm \hbar k \rangle_{i\,\,i\!}\langle0,0| = \hat{J}_-^\dag$, with the index $i$ labelling the atoms. Including the reverse processes, the many-body interaction Hamiltonian describing light scattering between pump field and cavity field reads (see Methods)
\begin{equation}
\frac{\hbar \lambda}{\sqrt{N}}(\hat{a}^\dag + \hat{a})(\hat{J}_+ + \hat{J}_-).
\end{equation}
This is exactly the interaction Hamiltonian of the Dicke model which describes $N$ two-level systems with transition frequency $\omega_0$ interacting with a bosonic field mode at frequency $\omega$. This system exhibits a quantum phase transition from a normal phase to a superradiant phase once the coupling strength $\lambda$ between atoms and light reaches the critical value of\cite{hepp1973, wang1973} $\lambda_\mathrm{cr} = \sqrt{\omega_0\omega}/2$. Our system realizes the Dicke Hamiltonian with $\omega = -\Delta_c + U_0 N/2$, $\omega_0 = 2\omega_r$ and $\lambda = \eta\sqrt{N}/2$. Correspondingly, the process of self-organization is equivalent to the Dicke quantum phase transition where both the cavity field and the atomic polarization $\langle \hat{J}_+ + \hat{J}_-\rangle = 2 \Theta$ acquire macroscopic occupations.

The experimental realization of the Dicke quantum phase transition is usually inhibited because the transition frequencies by far exceed the available dipole coupling strengths. Using optical Raman transitions instead brings the energy difference between the atomic modes from the optical scale to a much lower energy scale, which makes the phase transition experimentally accessible. A similar realization of an effective Dicke Hamiltonian has been theoretically considered using two balanced Raman channels between different electronic (instead of motional) states of an atomic ensemble interacting with an optical cavity and an external pump field\cite{dimer2007}. It is important to point out that these systems are externally driven and subject to cavity loss. Therefore they realize a dynamical version of the original Dicke quantum phase transition. However, the cavity output field offers the unique possibility to \emph{in situ} monitor the phase transition as well as to extract important properties of the system\cite{dimer2007}.

\section*{Experimental Description}

Our experimental setup has been described previously\cite{ottl2006, brennecke2007}. In brief, we prepare almost pure Bose-Einstein condensates of typically $10^5$ $^{87}\mathrm{Rb}$ atoms in a crossed-beam dipole trap which is centered inside an ultrahigh-finesse optical Fabry-Perot cavity. The atoms are prepared in the $|F, m_F\rangle =|1,-1\rangle$ hyperfine ground state, where $F$ denotes the total angular momentum and $m_F$ the magnetic quantum number. Perpendicular to the cavity axis the atoms are driven by a linearly polarized standing-wave laser beam whose wavelength $\lambda_p$ is red-detuned by $\unit[4.3]{nm}$ from the atomic $D_2$ line. The pump-atom detuning is more than five orders of magnitude larger than the atomic linewidth. This justifies that we neglect spontaneous scattering in our theoretical description, and consider only coherent scattering between the pump beam and a particular $\mathrm{TEM}_{00}$ cavity mode which is quasi-resonant with the pump laser frequency. The system operates in the regime of strong dispersive coupling\cite{klinner2006} where the maximum dispersive shift of the empty cavity resonance induced by all atoms, $NU_0$, exceeds the cavity decay rate $\kappa = 2\pi\times \unit[1.3]{MHz}$ by a factor of $6.5$.

The light leaking out of the optical resonator is detected with calibrated single-photon counting modules allowing us to \emph{in-situ} monitor the intracavity light intensity. In addition, we infer about the atomic momentum distribution from absorption imaging along the $y$-axis after a few milliseconds of free ballistic expansion of the atomic cloud.

\section*{Observing the Phase Transition}
\begin{figure}
\centering
\includegraphics{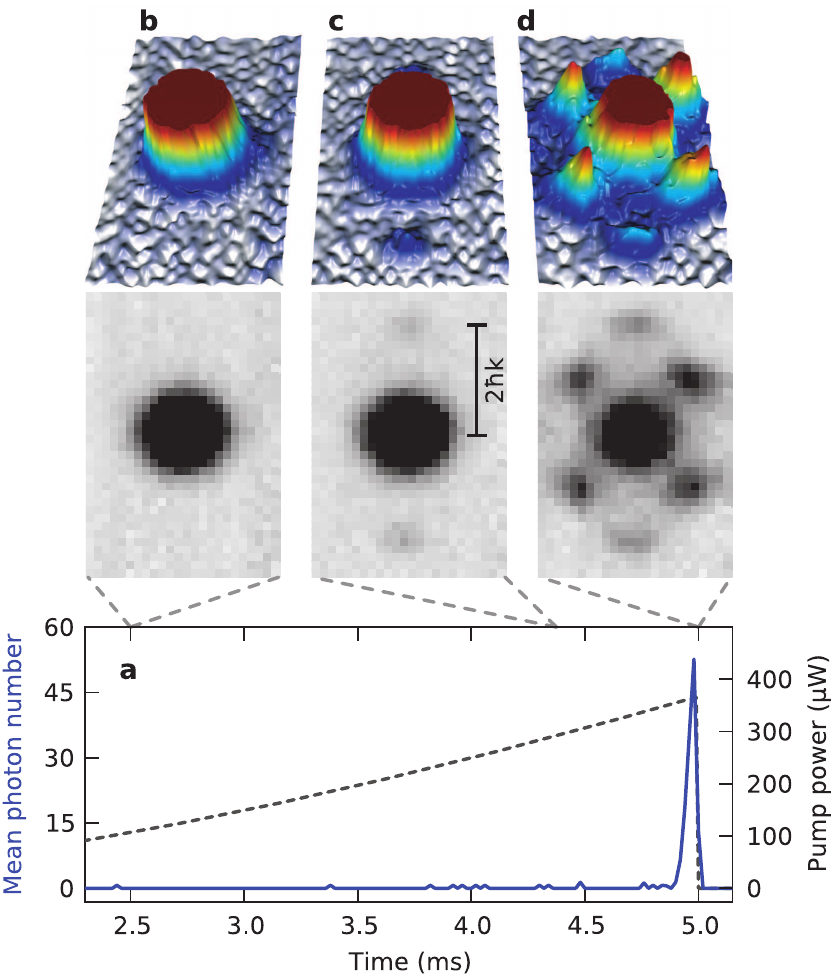}
\caption{Observation of the phase transition. \textbf{a}, The pump power (dashed) is gradually increased while monitoring the mean intracavity photon number (solid, binned over $\unit[20]{\mu s}$). After sudden release and subsequent ballistic expansion of $\unit[6]{ms}$, absorption images (clipped equally in atomic density) are taken for different pump powers corresponding to lattice depths of: \textbf{b}, $\unit[2.6]{E_{r}}$, \textbf{c}, $\unit[7.0]{E_{r}}$, \textbf{d}, $\unit[8.8]{E_{r}}$. Self-organization is manifested by an abrupt build-up of the cavity field accompanied by the formation of momentum components at $(p_{x},p_{z})=(\pm\hbar k,\pm\hbar k)$ (\textbf{d}). The weak momentum components at $(p_x,p_z)=(0,\pm 2\hbar k)$ (\textbf{c}) originate from loading the atoms into the 1D standing-wave potential of the pump laser. The pump-cavity detuning was $\Delta_{c}=\unit[-
2\pi \times 14.9(2)]{MHz}$ and the atom number $N=1.5(3)\times 10^5$.}
\end{figure}
To observe the onset of self-organization in the transversally pumped BEC, we gradually increase the pump power over time while monitoring the light leaking out of the cavity, see figure 1. As long as the pump power is kept below a threshold value no light is detected at the cavity output, and the expected momentum distribution of a condensate loaded into the shallow standing-wave potential of the pump field is observed (see Fig.~3b,c). Once the pump power reaches the critical value an abrupt build-up of the mean intracavity photon number marks the onset of self-organization (see Fig.~3a). Simultaneously, the atomic momentum distribution undergoes a striking change to show additional momentum components at $(p_x, p_z) = (\pm \hbar k, \pm \hbar k)$ (see Fig.~3d). This provides direct evidence for the acquired density modulation along one of the two sublattices of a checkerboard pattern associated with a non-zero order parameter $\Theta$.

Conceptually, the self-organized quantum gas can be regarded as a supersolid\cite{gopalakrishnan2009}, similar to those proposed for two-component systems\cite{buchler2003}. This requires the coexistence of non-trivial diagonal long-range order corresponding to a periodic density modulation, and off-diagonal long-range order associated with phase coherence. In our system the checkerboard structure of the density modulation is determined by the long-range cavity-mediated atom-atom interactions in a non-trivial way. This is because the arrangement of the atoms is restricted to two possible
checkerboard patterns which are intimately linked to the spontaneous breaking of the relative phase between pump and cavity field. In contrast, the spatial atomic structure in traditional optical lattice experiments is solely given by the externally applied light fields (see Methods). In addition, the off-diagonal long-range order of the Bose-Einstein condensate is not destroyed by the phase transition. The atomic coherence length extends over almost the full atomic ensemble, as we can deduce from the width of the higher-order momentum peaks in Fig.~3d.
\begin{figure}
\centering
\includegraphics{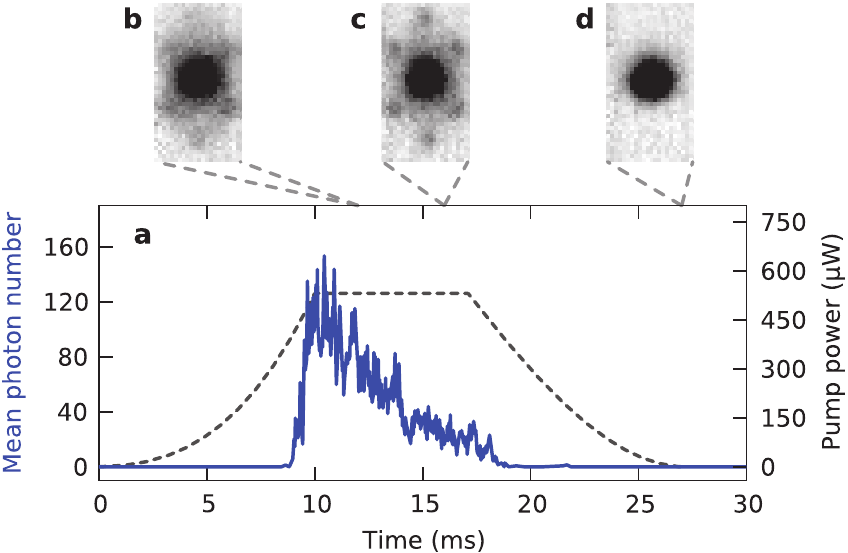}
\caption{Steady state in the self-organized phase. \textbf{a}, Pump power sequence (dashed) and recorded mean intracavity photon number (solid, binned over $\unit[20]{\mu s}$). After crossing the transition point at $\unit[9]{ms}$, the system reaches a steady state within the self-organized phase. The slow decrease in photon number is due to atom loss (see text). The short-time fluctuations are due to detection shot-noise. \textbf{b-d}, Absorption images are
taken after different times in the phase: (\textbf{b}) $\unit[3]{ms}$, (\textbf{c}) $\unit[7]{ms}$, and (\textbf{d}) after lowering the pump power again to zero. The pump-cavity detuning was $\Delta_c = -2\pi \times \unit[6.3(2)]{MHz}$ and the atom number $N=0.7(1)\times 10^5$.}
\end{figure}

After crossing the phase transition the system quickly reaches a steady state in the organized phase. As shown by a typical photon trace (see Fig.~4a), light is scattered into the cavity for up to $\unit[10]{ms}$ while the pump intensity is kept constant. This shows that the organized phase is stabilized by scattering induced light forces, which is in strict contrast to previous experiments observing (cavity-enhanced) superradiant light scattering\cite{inouye1999, slama2007} where a net transfer of momentum on the atomic cloud inhibited a steady state. The overall decrease of the mean cavity photon number for constant pump intensity (see Fig.~4a) is attributed to atom loss caused by residual spontaneous scattering at a rate of $\Gamma_\mathrm{sc} = \unit[3.7]{/s}$ and backaction-induced heating of the atoms\cite{murch2008}. Atom loss raises the critical pump power according to $P_\mathrm{cr} \propto N^{-1}$ which, close to the transition point, explains the observed reduction of the mean intracavity photon number. This was confirmed by entering the organized phase twice within one run and comparing the corresponding critical pump powers of self-organization. From absorption imaging we deduce an overall atom loss of $30\%$ for the pump-power sequence shown in Fig.~4a. Experimentally however, the atom-loss induced photon-number reduction can be compensated for by either steadily increasing the pump intensity or chirping the pump-cavity detuning.

From Fig.~4a we infer a maximum depth of the checkerboard lattice potential of $\unit[22]{E_r}$ which corresponds to single-site trapping frequencies of $\unit[19]{kHz}$ and $\unit[30]{kHz}$ along the $x$- and $z$-direction, respectively. Accordingly, the atoms are confined to an array of tubes which are oriented along the weakly confined $y$-direction and contain on average a few hundred atoms. Due to the strongly suppressed tunnelling rate between adjacent tubes separated by $\lambda_p/\sqrt{2}$ a dephasing of the different tubes is expected\cite{orzel2001}. This is directly observed via the reduced interference contrast in the absorption images reflecting that the supersolid phase evolved into a normal crystalline phase (see Fig.~4b). However, the phase coherence between the tubes is quickly restored when the mean intracavity photon number decreases and the lattice depth correspondingly lowers (see Fig.~4c). After ramping the pump intensity to zero, an almost pure BEC is retrieved (see Fig.~4d).

\section*{Mapping out the Phase Diagram}
\begin{figure*}
\centering
\includegraphics{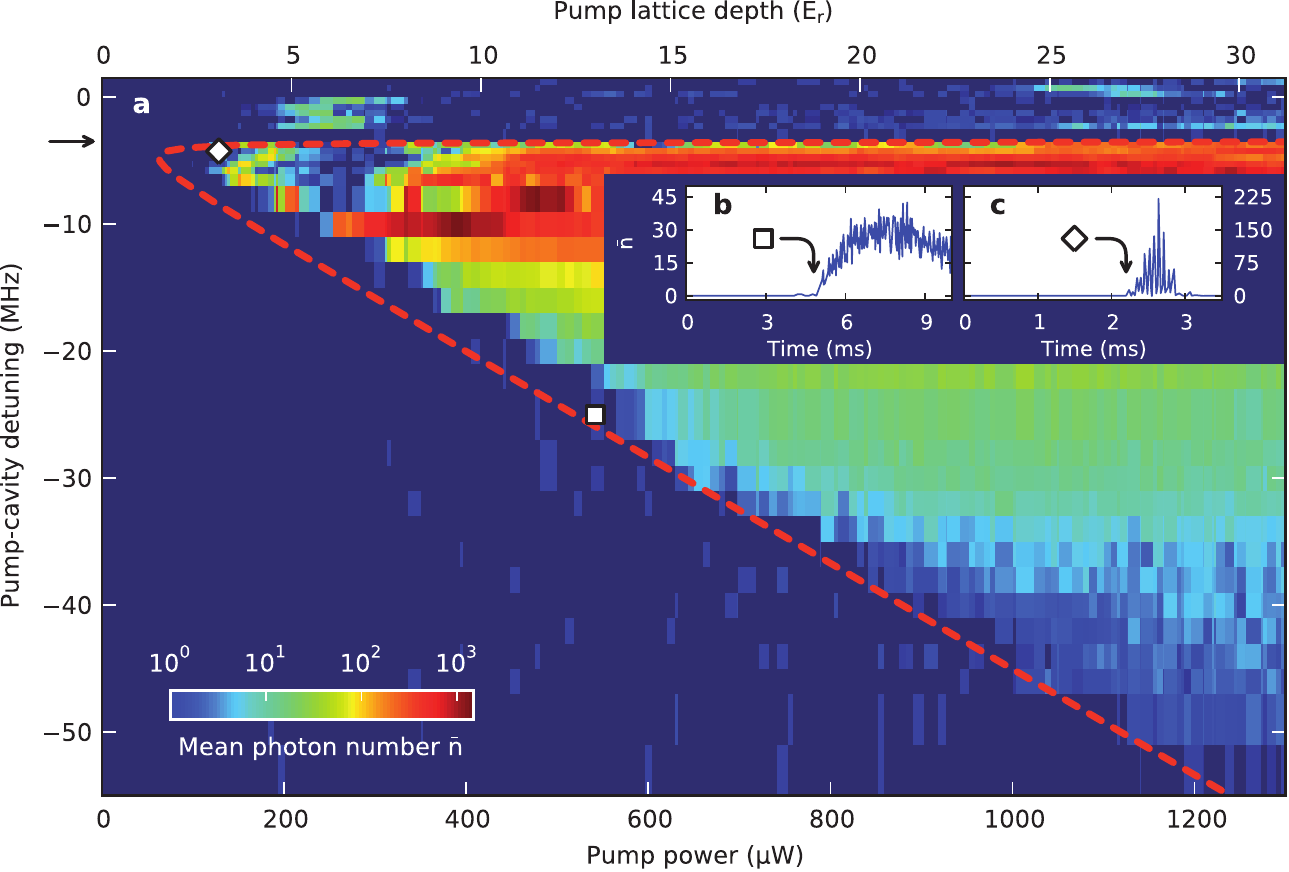}
\caption{Phase diagram. \textbf{a}, The pump power is increased to $\unit[1.3]{mW}$ over $\unit[10]{ms}$ for different values of the pump-cavity detuning $\Delta_{c}$. The recorded mean intracavity photon number $\bar{n}$ is displayed in color along the rescaled horizontal axis, showing pump power and corresponding pump lattice depth. A sharp phase boundary is observed over a wide range of the pump-cavity detuning $\Delta_c$, which is in very good agreement with a theoretical mean-field model (dashed curve). The dispersively shifted cavity resonance for the non-organized atom cloud is marked by the arrow. \textbf{b-c}, Typical traces showing the intracavity photon number for different pump-cavity detuning: (\textbf{b}) $\Delta_{c}=\unit[- 2 \pi\times 23.0(2)]{MHz}$, binned over $\unit[20]{\mu s}$, (\textbf{c}) $\Delta_{c}=\unit[- 2\pi \times 4.0(2)]{MHz}$, binned over $\unit[10]{\mu s}$. The atom number was $N= 1.0(2)\times 10^5$. In the detuning range $\unit[- 2\pi \times 7]{MHz} \geq \Delta_{c}\geq\unit[- 2\pi \times 21]{MHz}$ the pump power ramp was interrupted at $\unit[540]{\mu W}$. Therefore, no photon data was taken under the insets.}
\end{figure*}
From the analogy to the Dicke quantum phase transition we can deduce the dependence of the critical pump power on the pump-cavity detuning $\Delta_c$. To experimentally map out the phase boundary we gradually increase the pump power similar to Fig.~3a for different values of $\Delta_c$ (see Fig.~5b). The corresponding intracavity photon number traces are shown as a 2D color plot in Fig.~5a.

A sharp phase boundary is observed over a wide range of pump-cavity detuning $\Delta_c$. For large negative values of $\Delta_c$ the critical pump power $P_\mathrm{cr}\propto \lambda_\mathrm{cr}^2$ scales linearly with the effective cavity frequency $\omega = -\Delta_c+U_0 \mathcal{B}_0$, which agrees with the dependence expected from the Dicke model (see Methods). For $\omega < 0$, the critical coupling strength $\lambda_\mathrm{cr}$ has no real solution. Indeed, almost no light scattering is observed if the pump-cavity detuning is larger than the dispersively shifted cavity resonance at $U_0 \mathcal{B}_0 = -2\pi\times \unit[3.5]{MHz}$, where $\mathcal{B}_0$ denotes the spatial overlap between the cavity mode profile and the atomic density in the non-organized phase. As the pump-cavity detuning approaches the shifted cavity resonance from below, scattering into the cavity and the intracavity photon number increase.

We quantitatively compare our measurements with the phase boundary calculated in a mean-field description, including the external confinement of the atoms, the transverse pump and cavity mode profiles, and the collisional atom-atom interaction (see Methods). The agreement between measurements and theoretically expected phase boundary is excellent (see Fig.~5a, dashed curve).

The organization of the atoms on a checkerboard pattern not only affects the scattering rate between pump and cavity field, but also changes the spatial overlap\cite{brennecke2008} $\mathcal{B}$. This dynamically shifts the cavity resonance, which goes beyond the Dicke model (see Methods), and results in a frustrated system\cite{nagy2006} for $U_0 N > \Delta_c > U_0 \mathcal{B}_0$. Here the onset of self-organization brings the coupled atoms-cavity system into resonance with the pump laser, and the positive feedback which drives self-organization is interrupted (see Eq.~2). Experimentally this is observed in an oscillatory behavior of the system between the organized and the non-organized phase (see Fig.~5c).

\section*{Conclusions and Outlook}
We have experimentally realized a second-order dynamical quantum phase transition in a driven Bose-Einstein condensate coupled to the field of an ultrahigh-finesse optical cavity. At a critical driving strength the steady state realized by the system spontaneously breaks an Ising-type symmetry accompanied by self-organization of the superfluid atoms. We identify regimes where the emergent light-atom crystal is accompanied by phase coherence, and can thus be considered as a supersolid. The process of self-organization is shown to be equivalent to the Dicke quantum phase transition in an open system. We gain experimental access to the phase diagram of the Dicke model by observing the cavity output \emph{in situ}. In a very cold classical gas the corresponding phase boundary is predicted to scale with the temperature instead of the recoil energy\cite{asboth2005}, and the transition is driven by classical fluctuations in the atomic density instead of quantum fluctuations.

For the presented experiments the collective interaction $\lambda_\mathrm{cr}$ between the induced atomic dipoles and the cavity field approaches the order of the cavity decay rate $\kappa$, with a maximum ratio of $\lambda_\mathrm{cr}/\kappa = 0.2$. Reaching the regime where the Hamiltonian dynamics dominates the cavity losses offers possibilities to study the coherent dynamics of the Dicke model at the critical point which was shown theoretically to be dominated by macroscopic atom-field and atom-atom entanglement\cite{lambert2004,vidal2006,maschler2007}. Detecting the phase of the light leaving the resonator opens the opportunity to study spontaneous symmetry breaking induced by pure quantum fluctuations. Furthermore, recording the statistics of the scattered light may enable quantum non-demolition measurements and the preparation of exotic many-body states\cite{mekhov2007,mekhov2009a}.

\newpage
\section*{Methods}
\section*{Experimental Details}
We prepare almost pure $^{87}\mathrm{Rb}$ Bose-Einstein condensates in a crossed-beam dipole trap with trapping frequencies of $(\omega_x, \omega_y, \omega_z) = 2\pi\times\unit[(252,48,238)]{Hz}$, where $x$ denotes the cavity axis and $z$ the pump axis. For a typical atom number of $N = 10^5$ this results in condensate radii of $(R_x, R_y, R_z) = \unit[(3.2,16.6,3.3)]{\mu m}$ which were deduced in a mean-field approximation\cite{pitaevskii2003}. Experimentally, the position of the dipole trap is aligned to maximize the spatial overlap between the BEC and the cavity $\mathrm{TEM}_{00}$ mode which has a waist radius of $w_c = \unit[25]{\mu m}$. The cavity has a finesse of $3.4\times 10^5$. Its length of $\unit[178]{\mu m}$ is actively stabilized using a weak laser beam at $\unit[830]{nm}$ which is referenced onto the transverse pump laser\cite{ottl2006}. The intracavity stabilization light results in a weak lattice potential with a depth of less than $\unit[0.35]{E_r}$.

The pump laser beam has waist radii of $(w_x, w_y) = \unit[(29, 53)]{\mu m}$ at the position of the atoms. To accomplish optimal mode matching with the atomic cloud we use the same optical fiber for the pump light and the vertical beam of the crossed-beam dipole trap. The retro-reflected pump power is reduced by a factor of $0.6$ with respect to the incoming one due to clipping at the cavity mirrors and losses at the optical elements. All pump powers given in the text refer to the incoming one. The systematic uncertainties in determining the pump intensity at the position of the atoms is estimated to be $\unit[20]{\%}$. The pump light has a wavelength of $\lambda_{p}=\unit[784.5]{nm}$ and is linearly polarized along the $y$-axis (within an uncertainty of $\unit[5]{\%}$) to optimize scattering into the cavity mode. A weak magnetic field of $\unit[0.1]{G}$ pointing along the cavity axis provides a quantization axis for the atoms prepared in the $|F, m_F\rangle = |1, -1\rangle$ ground state. Accordingly, only $\sigma_+$ or $\sigma_-$ polarized photons can be scattered into the cavity mode. We observe the onset of self-organization always with $\sigma_+$ polarized cavity light since the corresponding atom-cavity coupling strength exceeds the one for $\sigma_-$ polarized light.

The light which leaks out of the cavity is monitored on two single-photon counting modules each of which is sensitive to one of the two different circular polarizations. In principle this allows to detect single intracavity photons with an efficiency of about $5\%$. However, for the experiments reported in this work the detection efficiency was reduced by a factor of $10$ in order to enlarge the dynamical range of our light detection (limited by the saturation effects of the photon counting modules). The systematic uncertainties in determining the intracavity photon number is estimated to be $\unit[25]{\%}$.

\section*{Mapping to the Dicke Hamiltonian}
The onset of self-organization is equivalent to a dynamical version of the normal to superradiant quantum phase transition of the Dicke model. This analogy is derived in a two-mode expansion of the atomic matter field, and allows to directly infer about properties of the transition into the organized phase. A one-dimensional analysis of the mapping to the Dicke model has been developed independently in Ref.~\cite{nagy2009}.

In the absence of collisional atom-atom interactions the many-body Hamiltonian describing the driven BEC--cavity system is given by
\begin{equation}
\hat{H} = \int \hat{\Psi}^\dag(x,z)\hat{H}_{(1)}\hat{\Psi}(x,z)\mathrm{d}x\,\mathrm{d}z
\end{equation}
where $\hat{\Psi}$ denotes the atomic field operator, and $\hat{H}_{(1)}$ is the single-particle Hamiltonian given in equation (1). In the non-organized phase the mean intracavity photon number vanishes and all atoms occupy the lowest-energy Bloch state $\psi_0$ of the 1D lattice Hamiltonian $\frac{\hat{p}_z^2}{2m} + V_0 \cos^2(k\hat{z})$. Scattering of photons between the pump field and the cavity mode couples the state $\psi_0$ to the state $\psi_1 \propto \psi_0 \cos(kx)\cos(kz)$ which carries additional $\hbar k$ momentum components along the $x$ and $z$ direction. In order to understand the onset of self-organization we expand the field operator $\hat{\Psi}$ in the reduced Hilbert space spanned by the modes $\psi_0$ and $\psi_1$.
Note that, for describing the deeply organized phase, higher-order momentum states have to be included in the description in order to account for atomic localization at the sites of the emergent checkerboard pattern.

After inserting the expansion $\hat{\Psi} = \psi_0 \hat{c}_0 + \psi_1 \hat{c}_1$ into the many-body Hamiltonian (see Eq.~4) we obtain up to a constant term
\begin{equation}
\hat{H}/\hbar = \omega_0 \hat{J}_z + \omega\hat{a}^\dag\hat{a} + \frac{ \lambda}{\sqrt{N}}(\hat{a}^\dag+\hat{a})(\hat{J}_+ + \hat{J}_-) + \frac{3}{4}U_0 \hat{c}_1^\dag\hat{c}_1\hat{a}^\dag\hat{a},
\end{equation}
with bosonic mode operators $\hat{c}_0$ and $\hat{c}_1$, and the total atom number $N = \hat{c}_0^\dag \hat{c}_0 + \hat{c}_1^\dag \hat{c}_1$. Here, the collective spin operators $\hat{J}_+ = \hat{c}_1^\dag\hat{c}_0 = \hat{J}^\dag_-$ and $\hat{J}_z = \frac{1}{2}(\hat{c}_1^\dag\hat{c}_1 - \hat{c}_0^\dag\hat{c}_0)$ were introduced. Apart from the last term, $\hat{H}$ is the Dicke Hamiltonian\cite{dimer2007} which describes the coupling between $N$ two-level systems with transition frequency $\omega_0 = 2\omega_r$ and a bosonic field mode with frequency $\omega = -\Delta_c + NU_0/2$. Their collective coupling strength is given by $\lambda = \sqrt{N}\eta/2$, which experimentally can be tuned by varying the pump laser power. The last term in equation (5) describes the dynamic (dispersive) shift of the cavity frequency, which is negligible in the close vicinity of the phase transition. Therefore, self-organization of the transversally pumped BEC--cavity system corresponds to the quantum phase transition of the Dicke model from a normal into a superradiant phase\cite{dimer2007}.

The Dicke Hamiltonian is invariant under the parity transformation\cite{lambert2004} $\hat{a}\rightarrow -\hat{a}$ and $\hat{J}_\pm \rightarrow -\hat{J}_\pm$. This symmetry is spontaneously broken by the process of self-organization corresponding to the atomic arrangement on the even or odd sites of a checkerboard pattern with $\langle \hat{J}_+ + \hat{J}_-\rangle = 2\Theta$ taking either positive or negative values. At the same time the relative phase between the pump and cavity field takes one of two possible values separated by $\pi$. This is in contrast to traditional optical lattice experiments where the phase between different laser beams determining their interference pattern is externally controlled\cite{greiner2001a}.

\section*{Derivation of the Phase Boundary in a Mean-Field Description}

To derive a quantitative expression for the critical pump intensity of self-organization, we perform a stability analysis of the compound BEC--cavity system in a mean-field description, following Ref.\cite{nagy2008}. For comparison with our experimental findings we take into account the external trapping potential, the transverse sizes of the cavity mode and the pump beam, as well as collisional atom-atom interactions. The system is described by the generalized Gross-Pitaevskii equation
\begin{align}
\Big(\frac{\mathbf{p}^2}{2m} &+ V_\mathrm{ext}(\mathbf{r}) + \hbar U_0|\alpha|^2\phi_c^2(\mathbf{r}) + \hbar\eta(\alpha+\alpha^*)\phi_c(\mathbf{r})\phi_p(\mathbf{r}) \notag\\&+ g|\psi|^2\Big)\psi(\mathbf{r},t) = \mu \psi(\mathbf{r},t)
\end{align}
where $\psi(\mathbf{r})$ denotes the condensate wave function (normalized to the total atom number $N$), and $\alpha$ denotes the coherent cavity field amplitude which was adiabatically eliminated according to:
\begin{equation}
\alpha = \frac{\eta\Theta}{\Delta_c - U_0\mathcal{B} + i\kappa}.
\end{equation}
The mode profiles of the cavity and the pump beam are given by $\phi_c(\mathbf{r}) = \cos(k x)e^{-\frac{y^2+z^2}{w_c^2}}$ and $\phi_p(\mathbf{r}) = \cos(k z)e^{-\frac{x^2}{w_x^2}-\frac{y^2}{w_y^2}}$, respectively. The external potential $V_\mathrm{ext}$ consists of the harmonic trapping potential $m (\omega_x^2 x^2 + \omega_y^2 y^2 + \omega_z^2 z^2)/2$ given by the crossed-beam dipole trap, and the lattice potential $V_0\phi_p^2(\mathbf{r})$ provided by the pump beam. The order parameter $\Theta = \langle\psi|\phi_c\phi_p|\psi\rangle$ and the bunching parameter $\mathcal{B} = \langle\psi|\phi_c^2|\psi\rangle$ are defined according to the main text. The collisional interaction strength is given by $g = \frac{4\pi \hbar^2 a}{m}$ with the s-wave scattering length $a$. The chemical potential of the condensate is denoted by $\mu$.

A defining condition for the critical two-photon Rabi frequency $\eta_\mathrm{cr}$ is obtained from a linear stability analysis of equation (6) around the non-organized phase $\psi_0$ with $\alpha = 0$. Starting with the two-mode ansatz $\psi = \psi_0(1+\epsilon\phi_c\phi_p)$ with $\epsilon \ll 1$, we carry out an infinitesimal propagation step into imaginary time in equation (6). This yields the following condition for the critical pump strength $\eta_\mathrm{cr}$ where the system exhibits a dynamical instability
\begin{equation}
\eta_\mathrm{cr}\sqrt{N_\mathrm{eff}} = \frac{1}{2}\sqrt{\frac{\tilde{\Delta}^2_c+\kappa^2}{- \tilde{\Delta}_c}}\sqrt{2\omega_r + 4 E_\mathrm{int}/\hbar}.
\end{equation}
Here, we introduced the effective number of maximally scattering atoms $N_\mathrm{eff} = \langle \psi_0|\phi_c^2 \phi_p^2|\psi_0\rangle$, and denoted the detuning between the pump frequency and the dispersively shifted cavity resonance by $\tilde{\Delta}_c = \Delta_c - U_0 \mathcal{B}_0$, with $\mathcal{B}_0 = \langle\psi_0|\phi_c^2|\psi_0\rangle$. The interaction energy per particle, given by $E_\mathrm{int} = \frac{g}{2N} \int|\psi_0|^4 d\mathbf{r}$, accounts for the mean-field shift of the free-particle dispersion relation.

Identifying $\omega = -\tilde{\Delta}_c$, $\omega_0 = 2\omega_r + 4 E_\mathrm{int}/\hbar$ and $\lambda_\mathrm{cr} = \eta_\mathrm{cr}\sqrt{N_\mathrm{eff}}$ our result agrees with the critical coupling strength $\lambda_\mathrm{cr}$ obtained in the Dicke model including cavity decay\cite{dimer2007}
\begin{equation}
\lambda_\mathrm{cr} = \frac{1}{2}\sqrt{\frac{\omega^2 + \kappa^2}{\omega}\omega_0}.
\end{equation}

The phase boundary shown in Fig.~5a (dashed curve) is obtained from equation (8) by approximating the condensate wave function $\psi_0$ by the Thomas-Fermi solution in the crossed-beam dipole trap\cite{pitaevskii2003}.

\begin{acknowledgments}
We thank G. Blatter, I. Carusotto, P. Domokos, A. Imamoglu, S. Leinss, R. Mottl, L. Pollet, H. Ritsch and M. Troyer for stimulating discussions. Financial funding from NAME-QUAM (European Union) and QSIT (ETH Z\"urich) is acknowledged. C.G. acknowledges ETH fellowship support.
\end{acknowledgments}

\end{document}